\documentclass[manuscript]{aastex}

\shortauthors{Roman et al.}

\begin{document}

\title{Regularization of the circular restricted three-body problem using 'similar' coordinate systems}
\shorttitle{Regularization of the 'similar' circular restricted three-body problem}
\author{R. Roman\altaffilmark{1} and I. Sz{\"u}cs-Csillik\altaffilmark{1}}
\affil{Astronomical Institute of Romanian Academy, Astronomical Observatory Cluj-Napoca,
               Str. Ciresilor No. 19, RO-400487 Cluj-Napoca, Romania}
\email{rdcroman@yahoo.com, iharka@gmail.com}

\begin{abstract}
The regularization of a new problem, namely the three-body problem, using 'similar' coordinate system is proposed.
For this purpose we use the relation of 'similarity', which has been introduced as an equivalence relation in a previous paper (see \cite{rom11}). First we write the Hamiltonian function, the equations of motion in canonical form, and then using a generating function, we obtain the transformed equations of motion. After the coordinates transformations, we introduce the fictitious time, to regularize the equations of motion. Explicit formulas are given for the regularization in the coordinate systems centered in the more massive and the less massive star of the binary system. The 'similar' polar angle's definition is introduced, in order to analyze the regularization's geometrical transformation. The effect of Levi-Civita's transformation is described in a geometrical manner. Using the resulted regularized equations, we analyze and compare these canonical equations numerically, for the Earth-Moon binary system.
\end{abstract}


\section{Introduction}

In a previous article (see \cite{rom11}), by introducing the "similarity" relation and applying it to the restricted three-body problem, the "similar" equations of motion were obtained.  These equations were connected with the classical equations of motion by some coordinate transformation relations (see equations (17) in \cite{rom11}). In this paper were also defined 'similar' parameters and physical quantities, and 'similar' initial conditions and some trajectories of the test particles into the physical $(x,S_1,y)$ and respectively $(x,S_2,y)$ planes were ploted.

Denoting $S_1$ and $S_2$ the components of the binary system (whose masses are $m_1$ and $m_2$), the equations of motion of the test particle (in the frame of the restricted three-body problem) in the coordinate system $(x,S_1,y,z)$ are (see equations (11)-(13) in \citep{rom11}):
\begin{equation}
\frac{d^2x}{dt^2}-2\frac{dy}{dt}=x-\frac{q}{1+q}-\frac{x}{(1+q)r_1^3}-\frac{q(x-1)}{(1+q)r_2^3}
\end{equation}

\begin{equation}
\frac{d^2y}{dt^2}+2\frac{dx}{dt}=y-\frac{y}{(1+q)r_1^3}-\frac{q\:y}{(1+q)r_2^3}
\end{equation}

\begin{equation}
\frac{d^2z}{dt^2}=-\frac{z}{(1+q)r_1^3}-\frac{q\:z}{(1+q)r_2^3}\;,
\end{equation}
where
\begin{equation}
r_1=\sqrt{x^2+y^2+z^2}\,\;,\;\;\;r_2=\sqrt{(x-1)^2+y^2+z^2}\;,\;\;\;\;\;q=\frac{m_2}{m_1}.
\end{equation}

In the 'similar' coordinate system $(x',S_2,y',z')$ the equations of motion of the test particle are (see equations (14)-(16) in \cite{rom11}):
\begin{equation}
\frac{d^2x'}{dt^2}+2\frac{dy'}{dt}=x'-\frac{q'}{1+q'}-\frac{x'}{(1+q')r_1^{'3}}-\frac{q'(x'-1)}{(1+q')r_2^{'3}}
\end{equation}

\begin{equation}
\frac{d^2y'}{dt^2}-2\frac{dx'}{dt}=y'-\frac{y'}{(1+q')r_1^{'3}}-\frac{q'\:y'}{(1+q')r_2^{'3}}
\end{equation}

\begin{equation}
\frac{d^2z'}{dt^2}=-\frac{z'}{(1+q')r_1^{'3}}-\frac{q'\:z'}{(1+q')r_2^{'3}}\;,
\end{equation}
where
\begin{equation}
r_1'=\sqrt{x'^2+y'^2+z'^2}\,\;,\;\;r_2'=\sqrt{(x'-1)^2+y'^2+z'^2}\;,\;\;\;q'=\frac{m_1}{m_2}\;.
\end{equation}

It can be easily verify that the equations of coordinate transformation are:
\begin{equation}
x'=1-x\;,\;\;y'=y\;,\;\;z'=z\,.
\end{equation}

One can observes that equations (1)-(3) and (5)-(7) have singularities in $r_1=0$, $r_2=0$, $r_1'=0$ and $r_2'=0$. These situations correspond to collision of the test particle whith $S_1$ or $S_2$ in a straight line. The collision is due to the nature of the Newtonian gravitational force ($F\propto \frac{1}{r^2}$). If the test particle approaches to one of the primaries very closely ($r\rightarrow 0$), then such an event produces large gravitational force ($F\rightarrow\infty$) and sharp bends of the orbit. The removing of these singularities can be done by regularization. (Remark: the purpose of regularization is to obtain regular equations of motion, no regular solutions.)

Euler seems to be the first (in 1765) to propose regularizing transformations when studying the motion of three bodies on a straight line (see \cite{sze75}). The regularization method has become popular in recent years (see \cite{jim11, cel11, wal06}) for long term studies of the motion of celestial bodies. These problems have a special merit in astronomy, because with their help we can studied more efficient the equations of motions with singularities. At the collision the equations of motion possess singularities. The problem of singularities plays an important role under computational, physical and conceptual aspects (see \cite{mio02, csi03}). The singularities occurring at collisions can be eliminated by the proper choice of the independent variable. The basic idea of regularization procedure is to compensate for the infinite increase of the velocity at collision. For this reason, a new independent variable, fictitious time, is adopted. The corresponding equations of motion are regularized by two transformations: the time transformation and the coordinate transformation. The most important part of the regularization is the time transformation, when a new fictitious time is used, in order to slow the motion near the singularities.

\section{The 'similar' canonical equations of motion}

The regularization can be local or global. If a local regularization is done, then the time and the coordinates  transformations eliminate only one of the two singularities. An example for the local regularization is the Birkhoff's transformation (see \cite{bir15}). The global regularization eliminates both singularities at once (see \cite{cas99, csi03}). Because our singularities are given in terms of $\frac{1}{r_1}$,  $\frac{1}{r_2}$, $\frac{1}{r'_1}$,  $\frac{1}{r'_2}$, in this paper a global regularization will be done.

In order to do this, we need to replace the cartesian equations
(1)-(3) and (5)-(7) with the corresponding canonical equations of
motion. The canonical coordinates are formed by generalized
coordinates $q_1, q_2, q_3$ and generalized momenta $p_1, p_2,
p_3$. The Hamiltonian, defined by the equation:
\begin{equation}
\mathcal{H}=\sum_{i=1}^{3}\dot{q_i}\;\frac{\partial \mathcal{L}}{\partial \dot{q_i}}-\mathcal{L}=\sum_{i=1}^{3}\dot{q_i}p_i-\mathcal{L}
\end{equation}
will becomes (see \cite{boc96} p. 266, for the generalized momenta, when the coordinates system rotates):
\begin{equation}
\mathcal{H}=\frac{1}{2}(p_1^2+p_2^2+p_3^2)+p_1 q_2- q_1 p_2 +\frac{q_1^2}{2}+\frac{q_2^2}{2}-\psi(q_1,q_2,q_3)\;.
\end{equation}
Here
\begin{equation}
\psi(q_1,q_2,q_3)=\frac{1}{2}\left[\left(q_1-\frac{q}{1+q}   \right)^2+q_2^2+\frac{2}{(1+q)r_1}  +\frac{2q}{(1+q)r_2} \right]\;,
\end{equation}
 with
\begin{equation}
r_1=\sqrt{q_1^2+q_2^2+q_3^2},\;\;\;\;r_2=\sqrt{(q_1-1)^2+q_2^2+q_3^2}\;.
\end{equation}
Here the generalized coordinates and the generalized momenta were:
\begin{equation}
q_1=x\;, \;\;q_2=y\;,\;\;q_3=z\;,\;\;p_1=\dot{q_1}-q_2\;\;\;p_2=\dot{q_2}+q_1\;,\;\;p_3=\dot{q_3}.
\end{equation}
Then, the canonical equations
\begin{equation}
\dot {q_i}=\frac{\partial \mathcal{H}}{\partial p_i}\;,\;\;\;\dot {p_i}=-\frac{\partial \mathcal{H}}{\partial q_i}\;, \;\;\;\;i\in\{1,2,3 \}
\end{equation}
have, in the $(q_1,S_1,q_2,q_3)$ coordinate system, the explicit
forme:
\begin{eqnarray}\label{eq1-3}
\frac{d q_1}{dt} &=& p_1+q_2 \\
\frac{d q_2}{dt} &=& p_2-q_1 \\
\frac{d q_3}{dt} &=& p_3 \\
\frac{d p_1}{dt} &=& p_2 - \frac{q}{1+q}-\frac{1}{1+q}\cdot
\frac{q_1}{r_1^3} - \frac{q}{1+q} \cdot \frac{q_1-1}{r_2^3} \\
\frac{d p_2}{dt} &=& -p_1 -\frac{1}{1+q}\cdot
\frac{q_2}{r_1^3} - \frac{q}{1+q} \cdot\frac{q_2}{r_2^3} \\
\frac{dp_3}{dt} &=& -\frac{1}{1+q} \cdot\frac{q_3}{r_1^3} - \frac{q}{1+q}\cdot \frac{q_3}{r_2^3}.
\end{eqnarray}
It is easy to verify that using the relations (14), the explicit
canonical equations become the cartesian equations (1)-(3).

In order to write the canonical equations in the 'similar'
coordinate system $(q_{1s},S_2,q_{2s},q_{3s})$, we have in view
the theoretical considerations from the article \citep{rom11}. The
index \textit{s} refers to 'similar' quantities. Then, the
'similar' Hamiltonian will be: (see \cite{boc96} p. 266):
\begin{equation}
\mathcal{H}_s=\frac{1}{2}(p_{1s}^2+p_{2s}^2+p_{3s}^2)-(p_{1s} q_{2s}- q_{1s} p_{2s}) +\frac{q_{1s}^2}{2}+\frac{q_{2s}^2}{2}-\psi_s(q_{1s},q_{2s},q_{3s})\;,
\end{equation}
where
\begin{equation}
\Psi_s(q_{1s},q_{2s},q_{3s})=\frac{1}{2}\left[\left(q_{1s}-\frac{q'}{1+q'}   \right)^2+q_{2s}^2+\frac{2}{(1+q')r_{1s}}  +\frac{2q'}{(1+q')r_{2s}} \right]\;,
\end{equation}
 with
\begin{equation}
r_{1s}=\sqrt{q_{1s}^2+q_{2s}^2+q_{3s}^2},\;\;\;\;r_{2s}=\sqrt{(q_{1s}-1)^2+q_{2s}^2+q_{3s}^2}\;.
\end{equation}
Here the generalized coordinates and the generalized momenta were:
\begin{equation}
q_{1s}=1-q_1\;, \;\;q_{2s}=q_2\;,\;\;q_{3s}=q_3\;,\;\;p_{1s}=-p_1\;\;\;p_{2s}=p_2-1\;,\;\;p_{3s}=p_3.
\end{equation}
Then, the canonical equations
\begin{equation}
\dot {q_{is}}=\frac{\partial {\mathcal{H}_s}}{\partial p_{is}}\;,\;\;\;\dot {p_{is}}=-\frac{\partial {\mathcal{H}_s}}{\partial q_{is}}\;, \;\;\;\;i\in\{1,2,3 \}
\end{equation}
have, in the $(q_{1s},S_2,q_{2s},q_{3s})$ coordinate system, the
explicit forme:
\begin{eqnarray}
\frac{d q_{1s}}{dt} &=&p_{1s}-q_{2s} \\
\frac{d q_{2s}}{dt} &=&p_{2s}+q_{1s} \\
\frac{d q_{3s}}{dt} &=& p_{3s}\\
\frac{d p_{1s}}{dt} &=&- p_{2s} - \frac{q'}{1+q'}-\frac{1}{1+q'}\cdot
\frac{q_{1s}}{r_{1s}^3} - \frac{q'}{1+q'} \cdot \frac{q_{1s}-1}{r_{2s}^3}\\
\frac{d p_{2s}}{dt} &=& p_{1s} -\frac{1}{1+q'}\cdot
\frac{q_{2s}}{r_{1s}^3} - \frac{q'}{1+q'} \cdot\frac{q_{2s}}{r_{2s}^3} \\
\frac{dp_{3s}}{dt} &=& -\frac{1}{1+q'} \cdot\frac{q_{3s}}{r_{1s}^3} - \frac{q'}{1+q'}\cdot \frac{q_{3s}}{r_{2s}^3}.
\end{eqnarray}
It is easy to verify that using the relations (27), (28), (29),
the explicit canonical equations (30), (31), (32), become the
cartesian equations (5)-(7).

\textit{Remark}: From equations (13) and (24) it is easy to observe that $r_1=r_{2s}$ and $r_2=r_{1s}$ (see also Figure 1 in \citep{rom11}).

\section{Coordinate transformation}
The equations of motion (19)-(21) and (30)-(32) have singularities
in  $r_{1}$ and $r_{2}$, respectively in $r_{1s}$ and $r_{2s}$. We
shall remove these singularities by regularization. Several
regularizing methods are known (see Stiefel et al. 1971). In this paper we shall use the Levi-Civita's
method, applied when the bodies are moving on a plane. The two steps performed in the process of regularization
of the restricted problem are the introduction of new coordinates
and the transformation of time. The combination of the coordinate
(dependent variable) transformation  and the time (independent
variable) transformation have an analytical importance and
increase  the numerical accuracy. For simplicity we shall consider
that the third body moves into the orbital plane.

\subsection{Case 1 - coordinate transformation in the coordinate system with origin in $S_1$}
For the regularization of the equations of motion in the $(q_1, S_1, q_2)$ coordinate system, we shall introduce new variables $Q_1$ and $Q_2$, conected with the coordinates $q_1$ and $q_2$ by the relations of Levi-Civita (see \cite{levi06}):
\begin{equation}
q_1=Q_1^2-Q_2^2\;,\;\;\;\;q_2=2Q_1Q_2\;,
\end{equation}
Let introduce the generating function $\mathcal{S}$ (see \cite{sti71}, p.196):
\begin{equation}\label{eq1-7}
\mathcal{S}=-p_1 f(Q_1, Q_2)-p_2 g(Q_1, Q_2)\;,
\end{equation}
a twice continuously differentiable function. Here $f$ and $g$ are
harmonic conjugated functions, with the property
\begin{eqnarray}
\frac{\partial f}{\partial Q_1} &=& \frac{\partial g}{\partial Q_2}\nonumber\\
\frac{\partial f}{\partial Q_2} &=& -\frac{\partial g}{\partial Q_1}\;\;.\nonumber
\end{eqnarray}
The generating equations are
\begin{equation}
q_i = -\frac{\partial \mathcal{S}}{\partial p_i}, \;\;\;\;P_i =-\frac{\partial \mathcal{S}}{\partial Q_i}\;,\;\;\;\;i\in \{1, 2 \}\;,
\end{equation}
with $P_1,\;P_2$ as new generalized momenta, or explicitly
\begin{eqnarray}\label{eq1-9}
q_1 &=& -\frac{\partial \mathcal{S}}{\partial p_1} =f(Q_1, Q_2) \nonumber\\
q_2 &=& -\frac{\partial \mathcal{S}}{\partial p_2}=g(Q_1, Q_2) \nonumber\\
P_1 &=&- \frac{\partial \mathcal{S}}{\partial Q_1}=p_1 \frac{\partial f}{\partial Q_1}+p_2 \frac{\partial g}{\partial Q_1}=p_1 a_{11}+p_2 a_{12} \nonumber\\
P_2 &=& -\frac{\partial \mathcal{S}}{\partial Q_2}=p_1 \frac{\partial f}{\partial Q_2}+p_2 \frac{\partial g}{\partial Q_2}=-p_1 a_{12}+p_2 a_{11}
\end{eqnarray}
where
\begin{eqnarray}
a_{11}&=&\frac{\partial f}{\partial Q_1} = \frac{\partial g}{\partial Q_2}\nonumber\\
a_{12}&=&-\frac{\partial f}{\partial Q_2} = \frac{\partial g}{\partial Q_1}\nonumber
\end{eqnarray}
Let introduce the following notation:
$${\bf A}= \left( \matrix{a_{11} & a_{12} \cr -a_{12} & a_{11}} \right), \;\; D=det {\bf A}= a_{11}^2+a_{12}^2\;,$$  
\begin{equation}
{\bf p}=\left( \matrix{p_1 \cr p_2 } \right), \;{\bf P}=\left( \matrix{P_1 \cr P_2} \right),\;\; {\bf P}=A\cdot  {\bf p};\;\;\textbf{p}=\frac{A^T}{D}\;\textbf{P},\;p_1^2+p_2^2=(P_1^2+P_2^2)/D\;,
\end{equation}
where $\textbf{A}^T$ represents the transpose of matrix $\textbf{A}$.
The new Hamiltonian with the generalized coordinates $Q_1$ and $Q_2$ and generalized momenta $P_1$ and $P_2$ is:
\begin{eqnarray}\label{eq1-11}
\mathcal{H}(Q_1,Q_2,P_1,P_2) &=& \frac{1}{2D} \left[ P_1^2+P_2^2 +P_1 \frac{\partial }{\partial Q_2} (f^2+g^2) - P_2 \frac{\partial }{\partial Q_1} (f^2+g^2) \right] + \frac{q}{1+q}f- \nonumber \\  &-& \frac{1}{1+q}\cdot \frac{1}{\overline{r}_1} - \frac{q}{1+q} \cdot \frac{1}{\overline{r}_2}-\frac{q^2}{2(1+q)^2}
\end{eqnarray}
where $\overline{r}_1=\sqrt{f^2+g^2}$,
$\overline{r}_2=\sqrt{(f-1)^2+g^2}$ and $D=4(Q_1^2+Q_2^2)$ and the
explicit canonical equations of motion in new variables become:
\begin{eqnarray}\label{eq1-12}
\frac{dQ_1}{dt}&=&\frac{1}{2D} \left[ 2P_1+\frac{\partial }{\partial Q_2} (f^2+g^2) \right] \nonumber\\
\frac{dQ_2}{dt}&=&\frac{1}{2D} \left[ 2P_2-\frac{\partial }{\partial Q_1} (f^2+g^2) \right ] \nonumber\\
\frac{dP_1}{dt}&=&- \frac{P_1}{2D} \cdot \frac{\partial }{\partial Q_1\partial Q_2} (f^2+g^2) +  \frac{P_2}{2D} \cdot \frac{\partial }{\partial Q_1 \partial Q_1} (f^2+g^2) - \frac{q}{1+q} \frac{\partial f}{\partial Q_1}+ \nonumber \\
&+&  \frac{1}{1+q}\cdot \frac{\partial}{ \partial Q_1} \left( \frac{1}{\overline{r}_1} \right) + \frac{q}{1+q} \cdot \frac{\partial}{ \partial Q_1} \left( \frac{1}{\overline{r}_2}  \right) \nonumber \\
\frac{dP_2}{dt}&=&- \frac{P_1}{2D} \cdot \frac{\partial }{\partial Q_2\partial Q_2} (f^2+g^2) +  \frac{P_2}{2D} \cdot \frac{\partial }{\partial Q_2 \partial Q_1} (f^2+g^2) - \frac{q}{1+q} \frac{\partial f}{\partial Q_2}+  \\
&+&  \frac{1}{1+q}\cdot \frac{\partial}{ \partial Q_2} \left( \frac{1}{\overline{r}_1} \right) + \frac{q}{1+q} \cdot \frac{\partial}{ \partial Q_2} \left( \frac{1}{\overline{r}_2}  \right)\nonumber
\end{eqnarray}
Using Levi-Civita's transformation $f=q_1=Q_1^2-Q_2^2\;,\;\;\;g=q_2=2Q_1Q_2$ (see relations (33)), the equations (39) becomes:
\begin{eqnarray}\label{eq1-14}
\frac{dQ_1}{dt}&=& \frac{P_1}{D} + \frac{Q_2}{2} \nonumber\\
\frac{dQ_2}{dt}&=& \frac{P_2}{D} - \frac{Q_1}{2} \nonumber\\
\frac{dP_1}{dt}&=& \frac{P_2}{2} - \frac{2qQ_1}{1+q}
-  \frac{2}{1+q} \frac{Q_1}{\overline{r}_1^2} - \frac{2q}{1+q} \frac{Q_1 (\overline{r}_1 -1)}{\overline{r}_2^3}+\frac{(P_1^2+P_2^2)Q_1}{4\overline{r}_1^2}  \\
\frac{dP_2}{dt}&=&-\frac{P_1}{2} + \frac{2qQ_2}{1+q}
-  \frac{2}{1+q} \frac{Q_2}{\overline{r}_1^2} - \frac{2q}{1+q} \frac{Q_2(\overline{r}_1+1)}{\overline{r}_2^3}+\frac{(P_1^2+P_2^2)Q_2}{4\overline{r}_1^2} \nonumber
\end{eqnarray}
where
$\overline{r}_1={Q_1^2+Q_2^2}$, $\;\;\;\overline{r}_2=\sqrt{(Q_1^2-Q_2^2-1)^2+4Q_1^2 Q_2^2}$, \\
with the new Hamiltonian
\begin{eqnarray}\label{eq1-15}
\mathcal{H}_{S1} &=& \frac{P_1^2+P_2^2}{8(Q_1^2+Q_2^2)} + \frac{1}{2} (P_1 Q_2 -P_2 Q_1)  + \frac{q}{1+q} (Q_1^2-Q_2^2)- \nonumber \\
&-& \frac{1}{1+q}\cdot \frac{1}{Q_1^2+Q_2^2} - \frac{q}{1+q} \cdot \frac{1}{\sqrt{(Q_1^2-Q_2^2-1)^2+4Q_1^2 Q_2^2}}-\frac{q^2}{2(1+q)^2}.
\end{eqnarray}

\subsection{Case 2 - coordinate transformation in the 'similar' coordinate system}

For the coordinate transformation in the $(q_{1s}, S_2, q_{2s})$ coordinate system, we introduce the generating function $\mathcal{S}_s$ in the plane $(q_{s1}, S_2, q_{s2})$, in the following form
\begin{equation}\label{eq1-16}
\mathcal{S}_s=-p_{s1} f_s(Q_{s1}, Q_{s2})-p_{s2} g_s(Q_{s1}, Q_{s2})\;,
\end{equation}
where $f_s$ and $g_s$ are harmonic conjugated functions.
The generating equations are
\begin{eqnarray}\label{eq1-17}
q_{si} &=& -\frac{\partial \mathcal{S}_s}{\partial p_{si}},  \nonumber\\
P_{si} &=& -\frac{\partial \mathcal{S}_s}{\partial Q_{si}}, \;\;\;i\in\{1, 2\}\;,
\end{eqnarray}
or explicitly
\begin{eqnarray}\label{eq1-18}
q_{s1} &=& -\frac{\partial \mathcal{S}_s}{\partial p_{s1}} =f_s(Q_{s1}, Q_{s2}) \nonumber\\
q_{s2} &=& -\frac{\partial \mathcal{S}_s}{\partial p_{s2}}=g_s(Q_{s1}, Q_{s2}) \nonumber\\
P_{s1} &=& -\frac{\partial \mathcal{S}_s}{\partial Q_{s1}}=p_{s1} \frac{\partial f_s}{\partial Q_{s1}}+p_{s2} \frac{\partial g_s}{\partial Q_{s1}}=p_{s1} b_{11}+p_{s2} b_{12} \\
P_{s2} &=& -\frac{\partial \mathcal{S}_s}{\partial Q_{s2}}=p_{s1} \frac{\partial f_s}{\partial Q_{s2}}+p_{s2} \frac{\partial g_s}{\partial Q_{s2}}=-p_{s1} b_{12}+p_{s2} b_{11}\nonumber
\end{eqnarray}
where
\begin{eqnarray}
b_{11}&=&\frac{\partial f_s}{\partial Q_{s1}} = \frac{\partial g_s}{\partial Q_{s2}}\nonumber\\
b_{12}&=&-\frac{\partial f_s}{\partial Q_{s2}} = \frac{\partial g_s}{\partial Q_{s1}}\nonumber
\end{eqnarray}
Let introduce the following notation, (\cite{sze67}, p. 373)
$${\bf B}= \left( \matrix{b_{11} & b_{12} \cr -b_{12} & b_{11}} \right),  D_s=det {\bf B}= b_{11}^2+b_{12}^2,$$
\begin{equation}
{\bf p_s}=\left( \matrix{p_{s1} \cr p_{s2} } \right), \;{\bf P_s}=\left( \matrix{P_{s1} \cr P_{s2}} \right),\; {\bf p_s}^2=\frac{1}{D_s} {\bf P_s}^2\;,p_{s1}^2+p_{s2}^2=(P_{s1}^2+P_{s2}^2)/D_s\;.
\end{equation}
The new Hamiltonian for the case 2 may be written
\begin{eqnarray}\label{eq1-20}
{\mathcal{H}_{S2}} &=& \frac{1}{2D_s} \left[ P_{s1}^2+P_{s2}^2 -P_{s1} \frac{\partial }{\partial Q_{s2}} (f_s^2+g_s^2) + P_{s2} \frac{\partial }{\partial Q_{s1}} (f_s^2+g_s^2) \right] + \frac{q'}{1+q'}f_s- \nonumber \\  &-& \frac{1}{1+q'}\cdot \frac{1}{\overline{r}_{s1}} - \frac{q'}{1+q'} \cdot \frac{1}{\overline{r}_{s2}}-\frac{q'^2}{2(1+q')^2}
\end{eqnarray}
where
$\overline{r}_{s1}=\sqrt{f_s^2+g_s^2}$, $\overline{r}_{s2}=\sqrt{(f_s-1)^2+g_s^2}$ and $D_s=4(Q_{s1}^2+Q_{s2}^2)$
and the Hamiltonian equations in new variables become
\begin{eqnarray}\label{eq1-21}
\frac{dQ_{s1}}{dt}&=&\frac{1}{2D_s} \left[ 2P_{s1}-\frac{\partial }{\partial Q_{s2}} (f_s^2+g_s^2) \right] \nonumber\\
\frac{dQ_{s2}}{dt}&=&\frac{1}{2D_s} \left[ 2P_{s2}+\frac{\partial }{\partial Q_{s1}} (f_s^2+g_s^2) \right ] \nonumber\\
\frac{dP_{s1}}{dt}&=& \frac{P_{s1}}{2D_s} \cdot \frac{\partial }{\partial Q_{s1}\partial Q_{s2}} (f_s^2+g_s^2) -  \frac{P_{s2}}{2D_s} \cdot \frac{\partial }{\partial Q_{s1} \partial Q_{s1}} (f_s^2+g_s^2) - \frac{q'}{1+q'} \frac{\partial f_s}{\partial Q_{s1}}+ \nonumber \\
&+& \frac{1}{1+q'}\cdot \frac{\partial}{ \partial Q_{s1}} \left( \frac{1}{\overline{r}_{s1}} \right) + \frac{q'}{1+q'} \cdot \frac{\partial}{ \partial Q_{s1}} \left( \frac{1}{\overline{r}_{s2}}  \right) \nonumber \\
\frac{dP_{s2}}{dt}&=& \frac{P_{s1}}{2D_s} \cdot \frac{\partial }{\partial Q_{s2}\partial Q_{s2}} (f_s^2+g_s^2) -  \frac{P_{s2}}{2D_s} \cdot \frac{\partial }{\partial Q_{s2} \partial Q_{s1}} (f_s^2+g_s^2) - \frac{q'}{1+q'} \frac{\partial f_s}{\partial Q_{s2}}+ \nonumber \\
&+& \frac{1}{1+q'}\cdot \frac{\partial}{ \partial Q_{s2}} \left( \frac{1}{\overline{r}_{s1}} \right) + \frac{q'}{1+q'} \cdot \frac{\partial}{ \partial Q_{s2}} \left( \frac{1}{\overline{r}_{s2}}  \right).
\end{eqnarray}
Because the singularity of the problem is given by the terms $1/r_{s1}$ and $1/r_{s2}$, we will made a global regularization using the Levi-Civita's transformation
\begin{equation}\label{eq1-22}
f_s=Q_{s1}^2-Q_{s2}^2, \hspace{1in} g_s=2Q_{s1} Q_{s2}
\end{equation}
The 'similar' Hamiltonian equations are given by
\begin{eqnarray}\label{eq1-23}
\frac{dQ_{s1}}{dt}&=& \frac{P_{s1}}{D_s} - \frac{Q_{s2}}{2} \nonumber\\
\frac{dQ_{s2}}{dt}&=& \frac{P_{s2}}{D_s} + \frac{Q_{s1}}{2} \nonumber\\
\frac{dP_{s1}}{dt}&=& -\frac{P_{s2}}{2} - \frac{2q'Q_{s1}}{1+q'}
-  \frac{2}{1+q'} \frac{Q_{s1}}{\overline{r}_{s1}^2} - \frac{2q'}{1+q'} \frac{Q_{s1} (\overline{r}_{s1} -1)}{\overline{r}_{s2}^3}+\frac{(P_{s1}^2+P_{s2}^2)Q_{s1}}{4\overline{r}_1^2} \nonumber \\
\frac{dP_{s2}}{dt}&=&\frac{P_{s1}}{2} + \frac{2q'Q_{s2}}{1+q'}
-  \frac{2}{1+q'} \frac{Q_{s2}}{\overline{r}_{s1}^2} - \frac{2q'}{1+q'} \frac{Q_{s2}(\overline{r}_{s1}+1)}{\overline{r}_{s2}^3}+\frac{(P_{s1}^2+P_{s2}^2)Q_{s2}}{4\overline{r}_1^2}.
\end{eqnarray}
where
$\overline{r}_{s1}={Q_{s1}^2+Q_{s2}^2}$, $\;\;\;\overline{r}_{s2}=\sqrt{(Q_{s1}^2-Q_{s2}^2-1)^2+4Q_{s1}^2 Q_{s2}^2}$,\\
with the new Hamiltonian
\begin{eqnarray}\label{eq1-24}
{\mathcal{H}}_{S2} &=& \frac{P_{s1}^2+P_{s2}^2}{8(Q_{s1}^2+Q_{s2}^2)} + \frac{1}{2} (P_{s2} Q_{s1}-P_{s1} Q_{s2})  + \frac{q'}{1+q'} (Q_{s1}^2-Q_{s2}^2)- \nonumber \\
&-& \frac{1}{1+q'}\cdot \frac{1}{Q_{s1}^2+Q_{s2}^2} - \frac{q'}{1+q'} \cdot \frac{1}{\sqrt{(Q_{s1}^2-Q_{s2}^2-1)^2+4Q_{s1}^2 Q_{s2}^2}}-\frac{q'^2}{2(1+q')^2}.
\end{eqnarray}

\section{Time transformation}

The transformation of the independent variable is necessary to achieve regularization. It is a slow-down treatment of the physical problem, a new time scale in which the motion slows down \citep{mik96}.

\subsection{Case 1 - time transformation in the coordinate system with origin in $S_1$}

To resolve the Hamiltonian equations (\ref{eq1-14}), we introduce the fictitious time $\tau$, (see \cite{sze67, wal72, wal82, erd04}), and making the time transformation
$\frac{dt}{d\tau}=\overline r_1^2 \overline r_2^3$, the new regular equations of motion are
\begin{eqnarray}\label{eq1-25}
\frac{dQ_1}{d\tau}&=& \frac{P_1 \overline r_1 \overline r_2^3}{4} + \frac{Q_2 \overline r_1^2 \overline r_2^3}{2} \nonumber\\
\frac{dQ_2}{d\tau}&=& \frac{P_2 \overline r_1 \overline r_2^3}{4} - \frac{Q_1 \overline r_1^2 \overline r_2^3}{2} \nonumber\\
\frac{dP_1}{d\tau}&=& \frac{P_2 \overline r_1^2 \overline r_2^3}{2} - \frac{2qQ_1}{1+q} \overline r_1^2 \overline r_2^3 \nonumber\\
&-&  \frac{2Q_1 \overline r_2^3}{1+q} - \frac{2q Q_1 (\overline{r}_1 -1)\overline r_1^2}{1+q}+\frac{(P_1^2+P_2^2)Q_1 \overline{r}_2^3}{4} \nonumber \\
\frac{dP_2}{d\tau}&=&-\frac{P_1 \overline r_1^2 \overline r_2^3}{2} + \frac{2qQ_2}{1+q} \overline r_1^2 \overline r_2^3 \nonumber\\
&-&   \frac{2Q_2 \overline r_2^3}{1+q} - \frac{2q Q_2 (\overline{r}_1 +1)\overline r_1^2}{1+q}+\frac{(P_1^2+P_2^2)Q_2 \overline{r}_2^3}{4}.
\end{eqnarray}
The explicit equations of motion  may be written
\begin{eqnarray}\label{eq1-26}
\frac{d^2 Q_1}{d\tau^2} &=& \frac{1}{4} \frac{dP_1}{d\tau} \overline r_1 \overline r_2^3 +\frac{1}{2} \frac{dQ_2}{d\tau} \overline r_1^2 \overline r_2^3+
\left( \frac{P_1}{2} +2 Q_2 \overline r_1 \right) \left( Q_1 \frac{d Q_1}{d \tau} + Q_2 \frac{d Q_2}{d \tau} \right) \overline r_2^3+\nonumber \\
&+&3\left( \frac{P_1}{2} +Q_2 \overline r_1 \right) \left( (Q_1^3+Q_1 Q_2^2-Q_1) \frac{d Q_1}{d \tau} + (Q_2^3+Q_1^2 Q_2+Q_2) \frac{d Q_2}{d \tau} \right) \overline r_1 \overline r_2 \nonumber \\
\frac{d^2 Q_2}{d\tau^2} &=& \frac{1}{4} \frac{dP_2}{d\tau} \overline r_1 \overline r_2^3 -\frac{1}{2} \frac{dQ_1}{d\tau} \overline r_1^2 \overline r_2^3+
\left( \frac{P_2}{2} -2 Q_1 \overline r_1 \right) \left( Q_1 \frac{d Q_1}{d \tau} + Q_2 \frac{d Q_2}{d \tau} \right) \overline r_2^3+\nonumber \\
&+&3\left( \frac{P_2}{2} -Q_1 \overline r_1 \right) \left( (Q_1^3+Q_1 Q_2^2-Q_1) \frac{d Q_1}{d \tau} + (Q_2^3+Q_1^2 Q_2+Q_2) \frac{d Q_2}{d \tau} \right) \overline r_1 \overline r_2
\end{eqnarray}

\textit{Remark}: It is easy to see that now, the equations of motion have no singularities.

For the application of the above problem in a binary system, we can obtain the solution in the form
\begin{eqnarray}\label{eq1-27}
q_1(t) &=& Q_1^2(t)-Q_2^2(t)\nonumber \\
q_2(t) &=& 2Q_1(t)Q_2(t)
\end{eqnarray}

\subsection{Case 2 - time transformation in the 'similar' coordinate system}

Introducing the fictitious time $\tau$ and making the time transformation $\frac{dt}{d\tau}=\overline r_{s1}^2 \overline r_{s2}^3$, the new regular equations of motion are obtained in the form
\begin{eqnarray}\label{eq1-28}
\frac{dQ_{s1}}{d\tau}&=& \frac{P_{s1} \overline r_{s1} \overline r_{s2}^3}{4} - \frac{Q_{s2} \overline r_{s1}^2 \overline r_{s2}^3}{2} \nonumber\\
\frac{dQ_{s2}}{d\tau}&=& \frac{P_{s2} \overline r_{s1} \overline r_{s2}^3}{4} +\frac{Q_{s1} \overline r_{s1}^2 \overline r_{s2}^3}{2} \nonumber\\
\frac{dP_{s1}}{d\tau}&=& -\frac{P_{s2} \overline r_{s1}^2 \overline r_{s2}^3}{2} - \frac{2q'Q_{s1}}{1+q'} \overline r_{s1}^2 \overline r_{s2}^3 \nonumber\\
&-&  \frac{2Q_{s1} \overline r_{s2}^3}{1+q'} - \frac{2q' Q_{s1} (\overline{r}_{s1} -1)\overline r_{s1}^2}{1+q'}+\frac{(P_{s1}^2+P_{s2}^2)Q_{s1} \overline{r}_{s2}^3}{4} \nonumber \\
\frac{dP_{s2}}{d\tau}&=&+\frac{P_{s1} \overline r_{s1}^2 \overline r_{s2}^3}{2} + \frac{2q'Q_{s2}}{1+q'} \overline r_{s1}^2 \overline r_{s2}^3 \nonumber\\
&-& \frac{2Q_{s2} \overline r_{s2}^3}{1+q'} - \frac{2q' Q_{s2} (\overline{r}_{s1} +1)\overline r_{s1}^2}{1+q'}+\frac{(P_{s1}^2+P_{s2}^2)Q_{s2} \overline{r}_{s2}^3}{4}.
\end{eqnarray}
The explicit equations of motion  are given by
\begin{eqnarray}\label{eq1-29}
\frac{d^2 Q_{s1}}{d\tau^2} = \frac{1}{4} \frac{dP_{s1}}{d\tau} \overline r_{s1} \overline r_{s2}^3 -\frac{1}{2} \frac{dQ_{s2}}{d\tau} \overline r_{s1}^2 \overline r_{s2}^3+
\left( \frac{P_{s1}}{2} - 2 Q_{s2} \overline r_{s1} \right) \left( Q_{s1} \frac{d Q_{s1}}{d \tau} + Q_{s2} \frac{d Q_{s2}}{d \tau} \right) \overline r_{s2}^3+\nonumber \\
+3\left( \frac{P_{s1}}{2} - Q_{s2} \overline r_{s1} \right) \left( (Q_{s1}^3+Q_{s1} Q_{s2}^2-Q_{s1}) \frac{d Q_{s1}}{d \tau} + (Q_{s2}^3+Q_{s1}^2 Q_{s2}+Q_{s2}) \frac{d Q_{s2}}{d \tau} \right) \overline r_{s1} \overline r_{s2} \nonumber \\
\frac{d^2 Q_{s2}}{d\tau^2} = \frac{1}{4} \frac{dP_{s2}}{d\tau} \overline r_{s1} \overline r_{s2}^3 + \frac{1}{2} \frac{dQ_{s1}}{d\tau} \overline r_{s1}^2 \overline r_{s2}^3+
\left( \frac{P_{s2}}{2} +2 Q_{s1} \overline r_{s1} \right) \left( Q_{s1} \frac{d Q_{s1}}{d \tau} + Q_{s2} \frac{d Q_{s2}}{d \tau} \right) \overline r_{s2}^3+\nonumber \\
+3\left( \frac{P_{s2}}{2} +Q_{s1} \overline r_{s1} \right) \left( (Q_{s1}^3+Q_{s1} Q_{s2}^2-Q_{s1}) \frac{d Q_{s1}}{d \tau} + (Q_{s2}^3+Q_{s1}^2 Q_{s2}+Q_{s2}) \frac{d Q_{s2}}{d \tau} \right) \overline r_{s1} \overline r_{s2}
\end{eqnarray}

\textit{Remark}: It is easy to see that the 'similar' equations of motion have no singularities.

For the application of the above problem in a binary system, we can obtain the solution in the form
\begin{eqnarray}\label{eq1-30}
q_{s1}(t) &=& Q_{s1}^2(t)-Q_{s2}^2(t)\nonumber \\
q_{s2}(t) &=& 2Q_{s1}(t)Q_{s2}(t)
\end{eqnarray}

\section{Numerical experiments}

For the numerical integration (Earth-Moon binary system), considering that the third body moves into the orbital plane (see \cite{kop78}), we used the initial values:
$$
q_{10}=0.6,\;\; q_{20}=0.4, \;\;p_{10}=0.1,\;\; p_{20}=0.6,\;\; t\in [0,2 \pi], \;\;q=0.0123 \;.
$$
For the numerical integration (Earth-Moon binary system) in the "similar" coordinate system we use the initial values (see eqs. (25)):
$$
q_{10s}=1.6,\;\; q_{20s}=0.4,\;\; p_{10s}=-0.1,\;\; p_{20s}=-1.6, \;\;t\in [0,2\pi],\;\; q'=81.30\;.
$$
For the numerical integration (Earth-Moon binary system) in the regularized coordinate system (equations (52)), we use the initial values (see also eq. (33)):
$$
Q_{10}=0.813,\;\; Q_{20}=0.246, \;\;P_{10}=0.458, \;\;P_{20}= 0.926, \;\;\tau \in [0,2\pi],\;\; q=0.0123 \;,
$$
and in the 'similar' regularized coordinate system (equations
(54)):
$$
Q_{10s}=1.275, \;\;Q_{20s}= 0.157,\;\; P_{10s}=-0.757,\;\; P_{20s}=-4.047, \;\;\tau \in [0, 2 \pi],\;\; q'=81.30\;.
$$
\subsection{Considerations on the initial conditions}
In Figure 1 we can compare the trajectories of the test particle in the coordinate systems with origin in $S_1$ (figures \textit{a}, \textit{c}, \textit{e}), and $S_2$ (figures \textit{b}, \textit{d}, \textit{f}). The point $P_1$ correspond to the initial conditions.

We consider the trajectories given in \citep{rom11} in figure 6 (in the coordinate systems $(x,S_1,y)$ and $(x',S_2,y')$) and we represented them in the coordinate systems $(q1,S_1,q2)$ and $(q1s,S_2,q2s)$ (see Figure 1 \textit{a} and \textit{b}). In this purpose we obtained the initial conditions as follows:
$$q_{10}=x_0=0.6\;;\;\;q_{20}=y_0=0.4\;;\;\;\dot{q}_{10}=v_{0x}=0.5\;;\;\;\dot{q}_{20}=v_{0y}=0\;,$$
and from eqs. (16)-(17):
$$p_{10}=\dot{q}_{10}-q_{20}=0.1\;;\;\;p_{20}=\dot{q}_{20}+q_{10}=0.6\;,$$
and
$$q_{10s}=x'_0=1.6\;;\;\;q_{20s}=y'_0=0.4\;;\;\;\dot{q}_{10s}=v'_{0x}=-0.5\;;\;\;\dot{q}_{20s}=v'_{0y}=0\;,$$
and from eqs. (27)-(28):
$$p_{10s}=\dot{q}_{10s}+q_{20s}=-0.1\;;\;\;p_{20s}=\dot{q}_{20s}-q_{10s}=-1.6\;.$$
In order to obtain the initial conditions, when we make the coordinate transformation, we solve the systems:

$ \;\;\;\;\;\;\;\;\;\left\{\begin{array}{rl}
       q_{10}&=Q_{10}^2-Q_{20}^2\\
       q_{20}&=2Q_{10}Q_{20}
          \end{array}\right.$,
$\;\;\; \left\{\begin{array}{rl}
       P_{10}&=2p_{10}Q_{10}+2p_{20}Q_{20}\\
       P_{20}&=-2p_{10}Q_{20}+2p_{20}Q_{10}
          \end{array}\right.$  \\(see eqs. (33) and (36)) for the trajectory in $(Q_1,S_1,Q_2)$ coordinate system (Figure 1c) and

 $ \;\;\;\;\;\;\;\;\;\left\{\begin{array}{rl}
       q_{10s}&=Q_{10s}^2-Q_{20s}^2\\
       q_{20s}&=2Q_{10s}Q_{20s}
          \end{array}\right.$,
$\;\;\; \left\{\begin{array}{rl}
       P_{10s}&=2p_{10s}Q_{10s}+2p_{20s}Q_{20s}\\
       P_{20s}&=-2p_{10s}Q_{20s}+2p_{20s}Q_{10s}
          \end{array}\right.$  \\(see eqs. (44) for the trajectory in $(Q_{1s},S_2,Q_{2s})$ coordinate system (Figure 1d).

Obviously, the initial conditions remain the same if we change the real time \textit{t} to the fictitious time $\tau$, but the motion is slowed. In Figure 1e and Figure 1f we represented the motion in real time \textit{t} with thin line and the slowed motion with thick line (corresponding to the same period of time).

\subsection{Considerations on the geometrical transformation}

Let us analyze the Figures 1a and 1c. For this purpose we consider
a point A($q_1,q_2$) on the graphic show in Figure 1a, and
B($Q_1,Q_2$) its corresponding point in Figure 1c. We have (see
Figure 2a and 2b):
$$ tan (\widehat{q_1S_1A})=\frac{q_2}{q_1}=\frac{2Q_1Q_2}{Q_1^2-Q_2^2}=\frac{2tan \widehat{BS_1Q_1}}{1-tan ^2\widehat{BS_1Q_1}}=tan (2\:\widehat{BS_1Q_1})$$
and it results: $\widehat{AS_1q_1}=2\:\widehat{BS_1Q_1}$. We used the counterclockwise directions for measuring the angles.

The Levi-Civita geometrical transformation originate in the conformal transformation (see \cite{boc96}, p.164):
$$z=q_1+i\:q_2=(Q_1+i\:Q_2)^2$$
where $(q_1,S_1,q_2)$ is the physical plane and $(Q_1,S_1,Q_2)$ is the parametric plane. From this relation we have:
$q_1=Q_1^2-Q_2^2$, $q_2=2\:Q_1Q_2$, and $\left|S_1A\right|=\left|S_1B\right|^2$. It means that the geometrical transformation squares the distances from the origin and doubled the polar angles.

If, having the trajectory in the physical plane, we want to draw
the trajectory into the parametric plane, we have to choose a
point $A_i$ on the trajectory in $(q_1,S_1,q_2)$ plane, measure
the angle $\widehat{A_iS_1q_1}$ and the distance $S_1A_i$, and
then draw a half-line $B'_iS_1$ in the $(Q_1S_1Q_2)$ plane, so as
$\widehat{A_iS_1q_1}=2\:\widehat{B'_iS_1Q_1}$. On this half-line,
we have to measure the distance $S_1B_i=\sqrt{(S_1A_i)}$, and
obtain the point $B_i$. Than we have to repeat the procedure for
$i=\overline{1,n}, \;\;n\in \mathbb{N}$. Of course the computer
will do this better and faster than we can do it, but the above
considerations help us to understand what it happened.

The vertex of the polar angles have to be centered into the more
massive star, so the angles $\widehat{q_1Aq_2}$ and
$\widehat{q_{1s}A_sq_{2s}}$ and respectively $\widehat{Q_1BQ_2}$
and $\widehat{Q_{1s}B_sQ_{2s}}$ are 'similar' polar angles. So, if
we intend to study the regularization of the circular restricted
three-body problem using 'similar' coordinate systems, we have to
add to 'similar' parameters postulated in section 3 in (Roman,
2011), \textit{the 'similar' polar angles}, measured between the abscissa
and the half-line passing through the center of the most massive
star and the test particle.

\begin{figure}
\begin{center}
 \includegraphics[height=0.8\textheight]{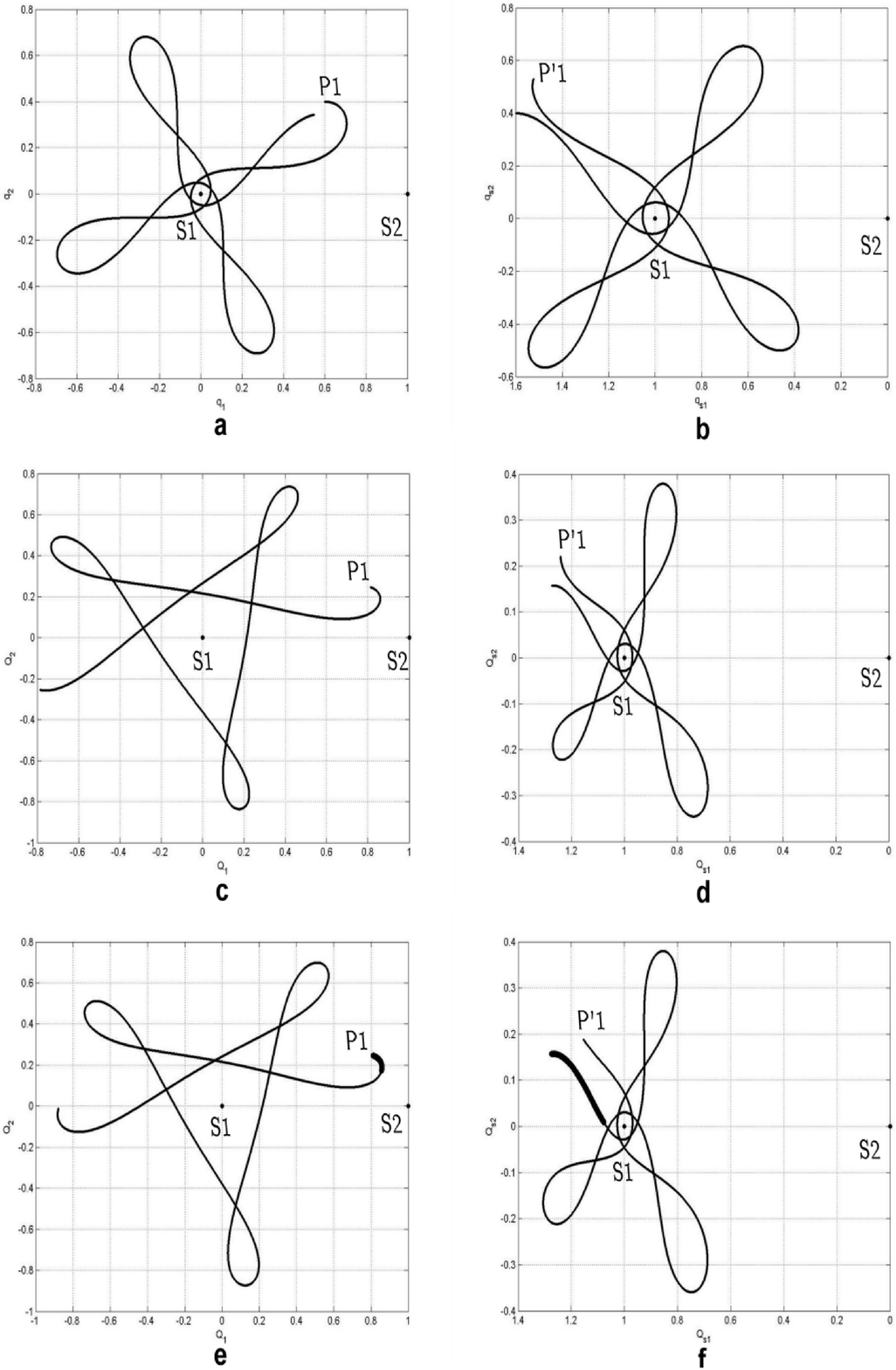}
 \caption{Trajectories in different coordinate systems with origin in S1 and S2}
\end{center}
\end{figure}

\begin{figure}
\begin{center}
  \includegraphics[height=0.3\textheight]{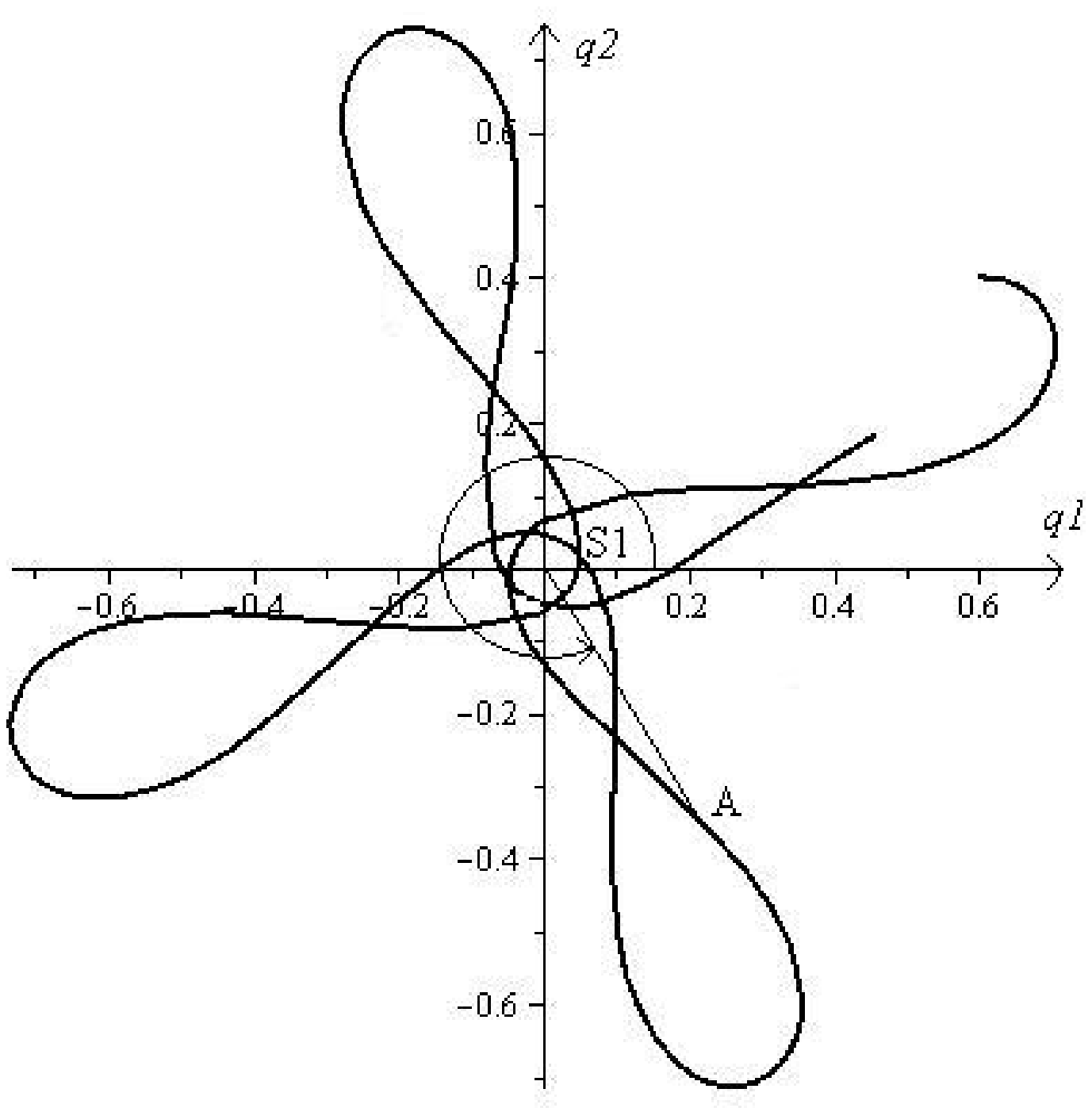}
  \hfill
  \includegraphics[height=0.3\textheight]{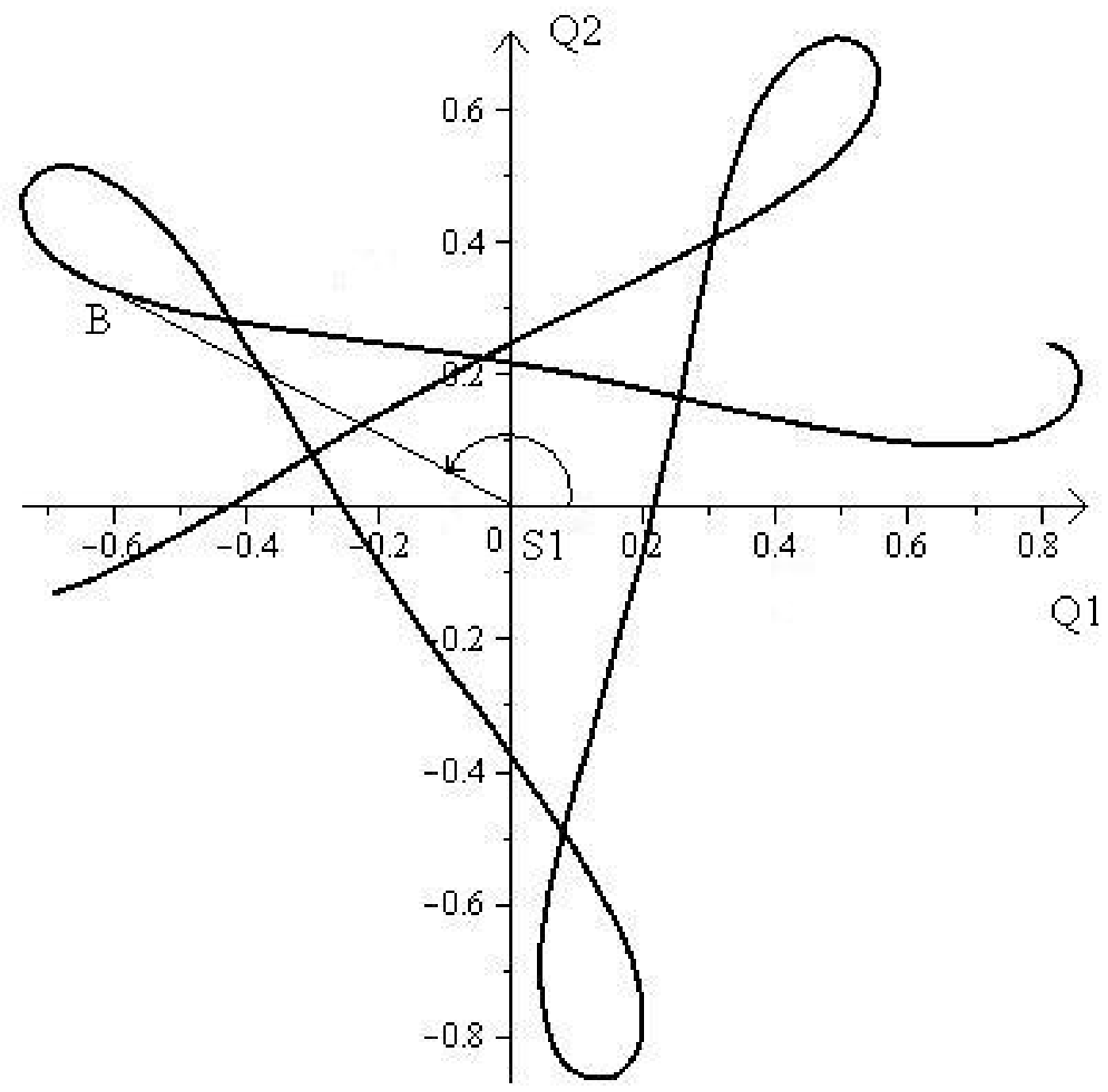}
  \caption{How to obtain the B point in (Q1,S1,Q2) plane from the A point from (q1,S1,q2) plane}
  \end{center}
\end{figure}

\begin{figure}
\begin{center}
  \includegraphics[height=0.3\textheight]{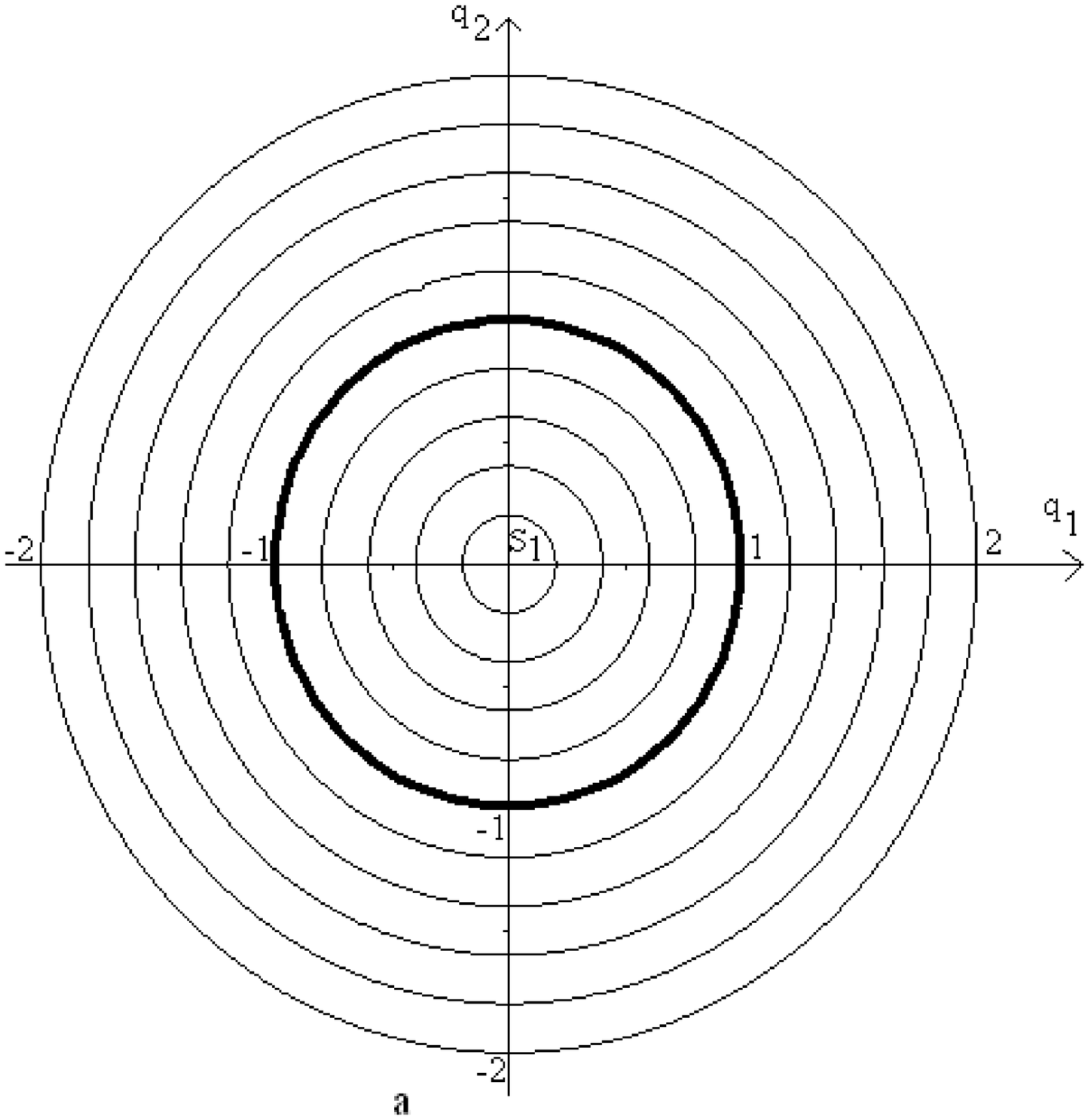}
  \includegraphics[height=0.3\textheight]{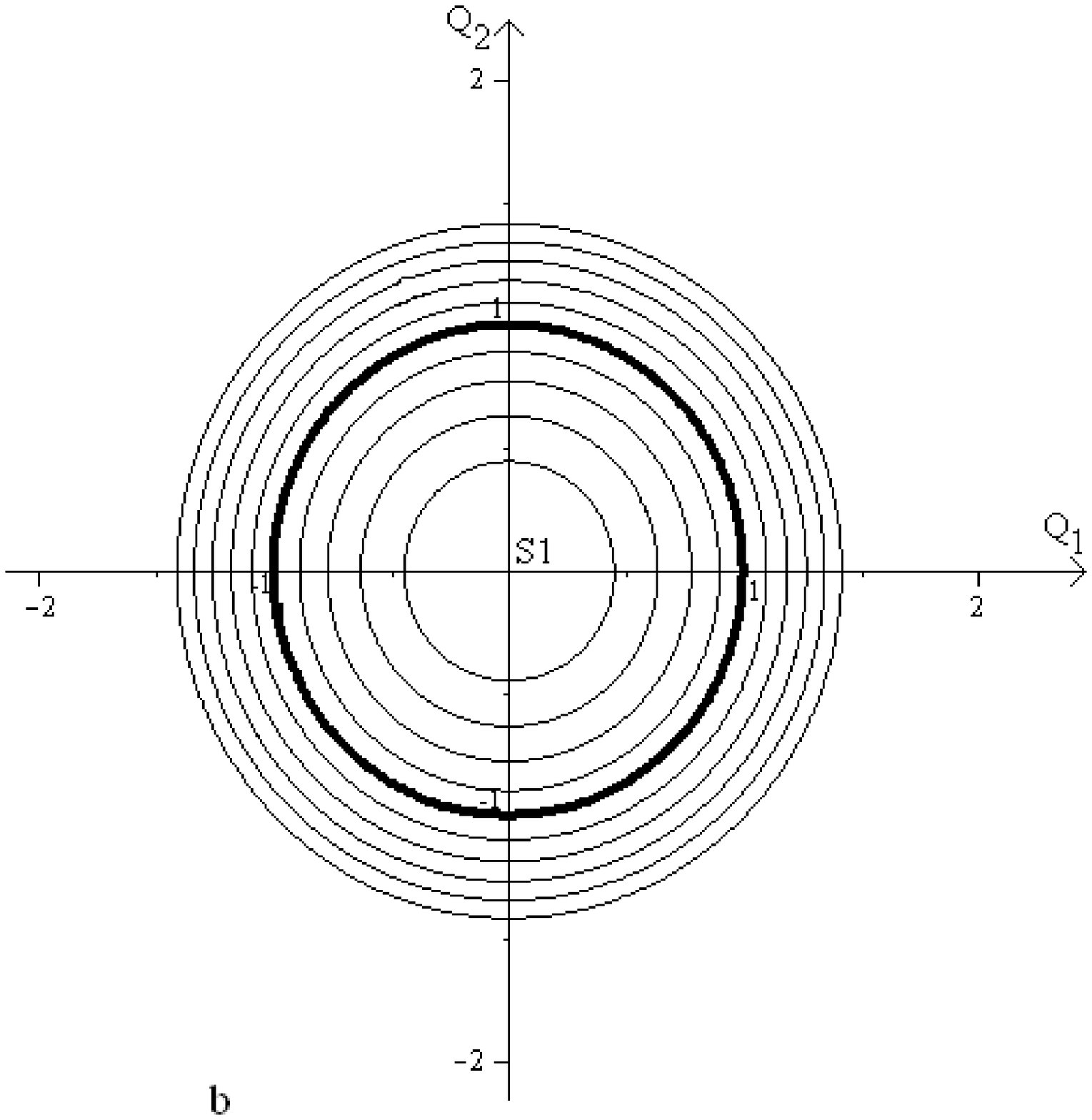}\\
  \includegraphics[height=0.3\textheight]{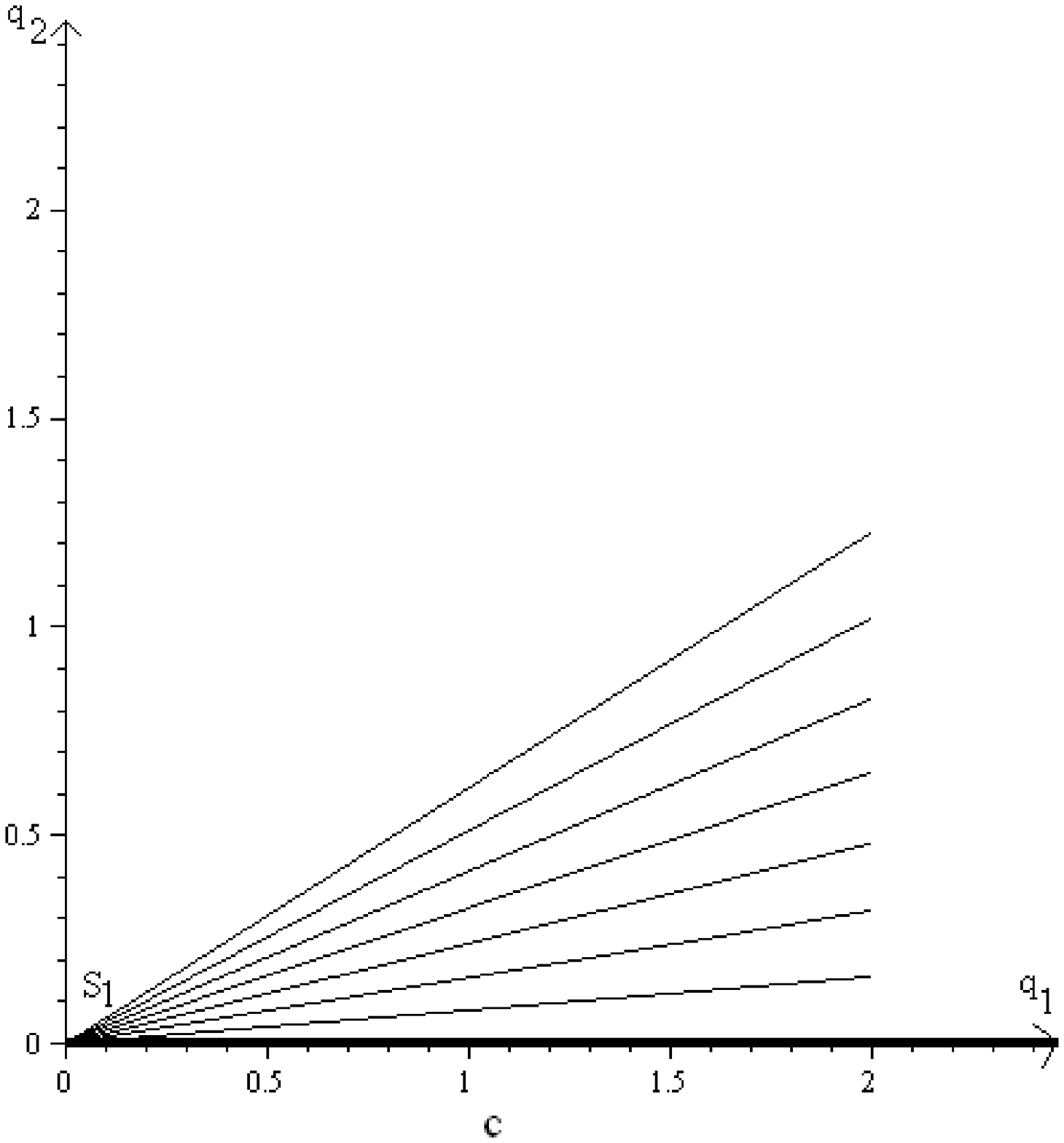}
  \includegraphics[height=0.3\textheight]{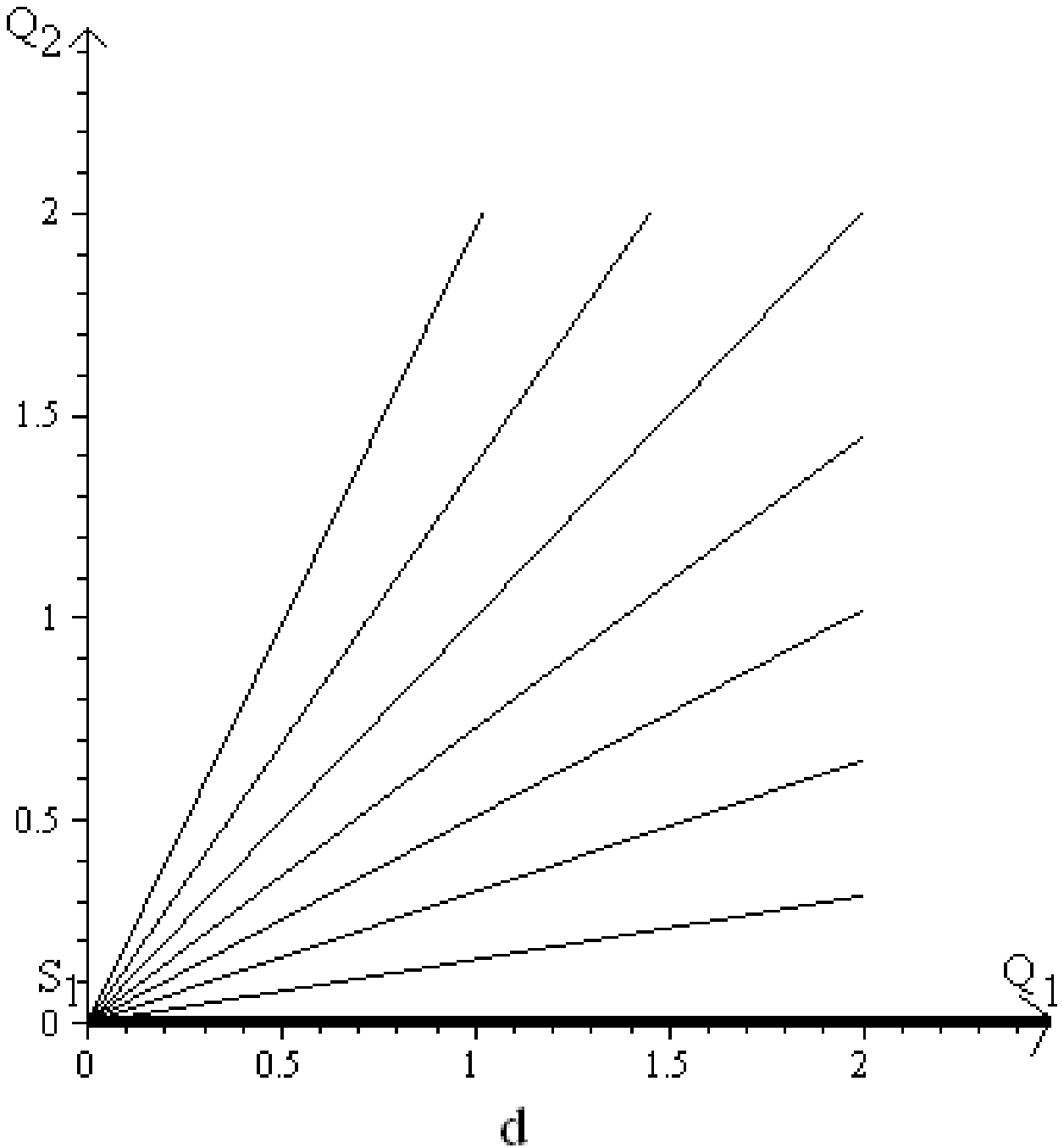}
  \caption{The role of the geometrical transformation in Levi-Civita's regularization}
  \end{center}
\end{figure}

\subsection{The effect of the Levi-Civita's regularization}
In order to see what is the effect of geometrical transformation,
let us analyze the graphics from Figure 3. In Figure 3a there are
represented some circles; their equations are: $q_1^2+q_2^2=r^2$,
where $r\in \{ 0.2; 0.4; 0.6; 0.8; 1; 1.2; 1.4; 1.6; 1.8; 2 \}$.
In figure 3b there are represented the circles having the
equations $Q_1^2+Q_2^2=u^2$, where $u=\sqrt{r}$, like geometrical
transformation of Levi-Civita's regularization postulated. One can
see that the circles in Figure 3b go away from the center and draw
near the circle having radius $u=1$. If in the center of the
circles there is a problem (a singularity), it can be easily
examined.

In Figure 3c and Figure 3d there are represented some half-lines,
having equations: $q_2=m\;q_1$, respectively $Q_2=n\;Q_1$, where
$m\in \{ 0; 0.2;0.3; 0.4; 0.5; 0.6; 0.7; 0.8 \}$, and $n=2\;m$,
like geometrical transformation of Levi-Civita's regularization
postulated. One can see that the half-lines in Figure 3d go away
from the abscissa's axis. If there is a problem (a singularity) on
the abscissa's axis, it can be easily examined.

There are only two points invariant with respect to the geometrical transformation of Levi-Civita's regularization: $S_1(0;0)$, and $S_2(1;0)$, respectively in the 'similar' coordinate systems $S_1(1;0)$, and $S_2(0;0)$. Then, the geometrical transformation go away the trajectory from the points where there are singularities.

In what concern the time transformation, as one can see in Figure 1e and 1f, the role of this transformation is to slow-down the motion of the test particle. With thin line is represented the trajectory of the test particle when the time integration is $2\;\pi$, and with thick line is represented the trajectory of the test particle when the time integration is $40\;\pi$. As one can see, after $40\;\pi$ we are still far away from the point where it is possible to have a singularity, if the coordinate system has the origin in $S_1$, but not so far away if the origin of the coordinate system is in $S_2$.

\section{Concluding remarks}

This paper continue the study of the relation of 'similarity', postulated in \citep{rom11}, by applying it to the Levi-Civita's regularization of the motion's equations of the test particle, in the circular, restricted three-body problem. Many papers in the last decade have studied the restricted three-body system in a phase space. During these studies, difficulties have arisen when the system approaches a close encounter.

Using the regularization method in the 'similar' coordinates system, we give explicitly equations of motion for the test particle. We study numerically the regular equations of motion, we written in canonical form, and obtained that the integrator using regularized equations of motion are more efficient. The 'similar' Hamiltonian (see eq. (22)) give us the 'similar' canonical equations (27)-(32), which have some different signes than the canonical equation (16)-(21). The coordinate transformation used in the Levi-Civita's regularization create a new form of the 'similar' Hamiltonian's equations (eqs. (49)). Finally, the time transformation used in the Levi-Civita's regularization gives us the regularized equations of motion (51) in the coordinate system with origin in $S_1$, and (54) in the 'similar' coordinate system.

In order to explain the shape of the trajectories in a concrete example, the 'similar' polar angle is introduced.

Our method may provide new directions for studies of circular restricted three-body integration using similar coordinate systems. It is an important tool for developing efficient numerical algorithms.

\end{document}